\documentclass[twoside]{article}
\usepackage{fleqn,espcrc2}

\usepackage{graphicx}

\newcommand{\epem}   {\ensuremath{\mathrm{e^+e^-}}}

\newcommand{\oaa}    {\ensuremath{\mathcal{O}(\alpha_s^2)}}
\newcommand{\bt}     {\ensuremath{B_T}}
\newcommand{\bw}     {\ensuremath{B_W}}

\newcommand{\thr}    {\ensuremath{1-T}}
\newcommand{\cp}     {\ensuremath{C}}
\newcommand{\cf}     {\ensuremath{C_F}}
\newcommand{\ca}     {\ensuremath{C_A}}

\newcommand{\nf}     {\ensuremath{n_f}}
\newcommand{\anull}  {\ensuremath{\alpha_0}}

\newcommand{\as}     {\ensuremath{\alpha_s}}
\newcommand{\asmu}   {\ensuremath{\alpha_s(\mu)}}
\newcommand{\mui}    {\ensuremath{\mu_I}}
\newcommand{\mil}    {\ensuremath{{\cal M}}}
\newcommand{\roots}  {\ensuremath{\sqrt{s}}}
\newcommand{\mz}     {\ensuremath{M_{\mathrm{Z^0}}}}
\newcommand{\asmz}   {\ensuremath{\alpha_s(M_{\mathrm{Z^0}})}}

\title{ Power correction analyses in \epem\ annihilation }

\author{ S. Kluth\address{ Max-Planck-Institut f\"ur Physik,\\
         F\"ohringer Ring 6, D-80805 Munich, Germany } }

\begin{document}

\begin{abstract}
The current status of theoretical work and experimental analyses on power
corrections in QCD for \epem\ annihilation will be reviewed. Measurements
of the number of active quark flavours \nf\ and the QCD colour factors \ca\
and \cf\ derived from QCD fits to event shape distributions at lower
energies and LEP energies will be presented. The fits are based on
\oaa+NLLA QCD predictions with power corrections to model
hadronisation.
\vspace{1pc}
\end{abstract}

\maketitle

\section{INTRODUCTION}

QCD studies always have to consider how the observed final states
consisting of hadrons, e.g. from the process $\epem\rightarrow$ hadrons,
map on QCD predictions in terms of partons, the quarks and gluons. The
transition from the partons of the hard process to the observed
hadrons, commonly referred to as hadronisation, is a fundamental
property of QCD. 

However, there is no fundamental QCD based theory of the transition
from partons to hadrons. The most successful description of the
hadronisation process in \epem\ annhilation has been achieved with
Monte Carlo simulations based on a parton shower followed by formation
of hadrons. Recently, analytic models of the hadronisation have
become available which predict how hadronisation effects of QCD
observables, e.g. event shapes, scale as a power of the energy scale
of the hard process. These models are often referred to as power
corrections.

In this report we will briefly review some of the analytic models and
present comparisons with \epem\ annihilation data. We will also show a
more general test of the consistency of power corrections with the
gauge structure of QCD.

\section{PREDICTIONS}

We summarise three related approaches to predict the power correction
of event shape observables. In all cases the form of the dependence on
the scale of the hard process is extracted from an analysis of
infrared renormalon singularities. The differences between the
approaches come from the regularisation of the infrared singularities
in order to obtain meaningful predictions.

\subsection{ DMW Model }

The model of Dokshitzer, Marchesini and Webber
(DMW)~\cite{dokshitzer95a,beneke95a} makes the assumption that
evolution of \as\ to energies below the Landau pole is possible but
the form of \asmu\ is a priori unknown. A non-perturbative parameter
\anull\ is introduced as the 0th moment over \asmu:
\begin{equation}
  \anull= \frac{1}{\mui}\int_0^{\mui}\as(k)dk\;\;\;.
\end{equation}
The quantity \mui\ is the infrared matching scale where the
non-perturbative and perturbative evolution of \as\ are merged,
usually $\mui=2$~GeV.

For some event shape observables the power corrections have been
calculated up to two loops~\cite{dokshitzer98b,dokshitzer99a}. The
result is that hadronisation is described by a shift of the
perturbative prediction $dR_{PT}/dy$ inversely proportional to the
hard scale $\roots=Q$:
\begin{equation}
  \frac{dR}{dy}= \frac{dR_{PT}}{dy}(y-PD_y)\;\;\;.
\label{equ_dmw}
\end{equation}
The factor $P\sim\mil\mui\anull(\mui)/Q$ is
universal~\cite{dokshitzer98b} and includes two loop corrections in
the quantity
$\mil=1+(1.575\ca-0.104\nf)/\beta_0$~\cite{dokshitzer99b}. The factor
$D_y$ contains all dependencies on the specific event shape observable
in question. The prediction for the 1st moment from equation
(\ref{equ_dmw}) is $\langle y\rangle= \langle y\rangle_{PT} + PD_y$. 
We note that the DMW model predicts that \anull\ does not
depend on the observable. This property of universality can be tested
experimentally.

\subsection{ Single dressed gluon approximation }
\label{sec_gardi}

A theoretical analysis of power corrections of the 1st and higher
moments of the thrust distribution was presented
in~\cite{gardi99a,gardi00a}. A different approach compared to the DMW
model is taken by adding a resummation based on the so-called single
dressed gluon approximation (SDG) to the standard perturbative QCD
predictions in \oaa. The SDG regularises the infrared renormalon
singularities. 

For the first moment of thrust a power correction proportional to
$1/Q$ is found, in agreement with the DMW model. The additional
resummation leads to a substantial reduction in the relative magnitude
of the power correction. It should be noted that power corrections
calculated in this approach cannot be directly compared to the DMW
model~\cite{dokshitzer99b}.

For the 2nd and 3rd moment of thrust the power corrections are
predicted to be of the form $1/Q^3$. However, other terms of the form
$1/Q^2$ and $\as/Q$ are also predicted to exist but cannot be
quantified.

\subsection{ Shape functions }

This approach is based on factorisation of soft and hard processes in
event shape distributions~\cite{korchemsky99a,korchemsky00a}. The
Ansatz is a convolution of the resummed perturbative QCD calculation
(\oaa+NLLA) $R_{PT}(y)$ for the hard process with a so-called shape
function $f(\epsilon)$ for the non-pertubative or soft processes. A
factorisation scale $\mu$ separates contributions from gluons with
$k_t<\mu$ taken account of by $f(\epsilon)$ from those with $k_t>\mu$
included in $R_{PT}(y)$. For a cumulative event shape distribution
$R(y)$ at cms energy $Q$ one has:
\begin{equation}
R(y)=
\int_0^{yQ}f(\epsilon,\mu)R_{PT}(y-\frac{\epsilon}{Q},\mu)d\epsilon\;\;\;. 
\end{equation}
It is shown that the shape function $f(\epsilon)$ resums all power
corrections of the form $1/(yQ)^n$ and that it is related to the
energy flow in the event. 

The description of non-perturbative effects by a shift $\sim 1/Q$ of
the perturbative QCD prediction in the DMW model is reproduced as the
special case where $f(\epsilon)$ is a delta function. For the 2nd
moments of thrust, heavy jet mass and C-parameter a prediction for the
power correction of the form $\langle
y\rangle_{PT}\lambda_1/Q+\lambda_2/Q^2$ is given, consistent with
section~\ref{sec_gardi}. Comparison of the predictions with
distributions, 1st and 2nd moments of thrust, heavy jet mass and
C-parameter from 35~GeV to 189~GeV shows reasonable agreement with the
data.

\section{ STANDARD ANALYSES }

All experimental analyses study power corrections in the DMW
model. For moments the perturbative QCD prediction is \oaa\ while for
distributions the power correction terms are combined with \oaa+NLLA
calculations.

\subsection{ First and second moments }

Studies of power corrections using the 1st moments of the event shape
observables thrust, heavy jet mass, total and wide jet broadening and
C-parameter have been performed by several collaborations. The LEP
experiments ALEPH~\cite{alephlep2data1},
DELPHI~\cite{delphilep2data1,delphilep2data2} and
L3~\cite{l3lep2data1,l3lep2data2} used data recorded at cms energies
up to 205~GeV while the reanalysis of data from the PETRA experiment
JADE covers in addition the range 35~GeV to 44~GeV~\cite{jadenewas,jadec}.

\begin{figure}[!htb]
\vspace{9pt}
\includegraphics[width=\columnwidth]{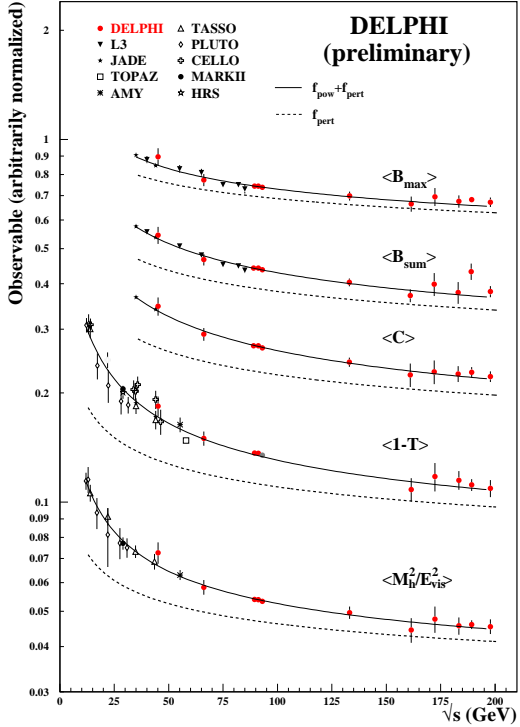}
\caption{First moments of event shapes from DELPHI~\cite{delphilep2data2}.}
\label{fig_delphi}
\end{figure}

Figure~\ref{fig_delphi} presents the DELPHI high energy data together
with data from other experiments at $\roots<\mz$. The fit of the \oaa\
perturbative QCD prediction combined with a power correction describes
the data well.

\begin{figure}[!htb]
\vspace{9pt}
\includegraphics[width=\columnwidth]{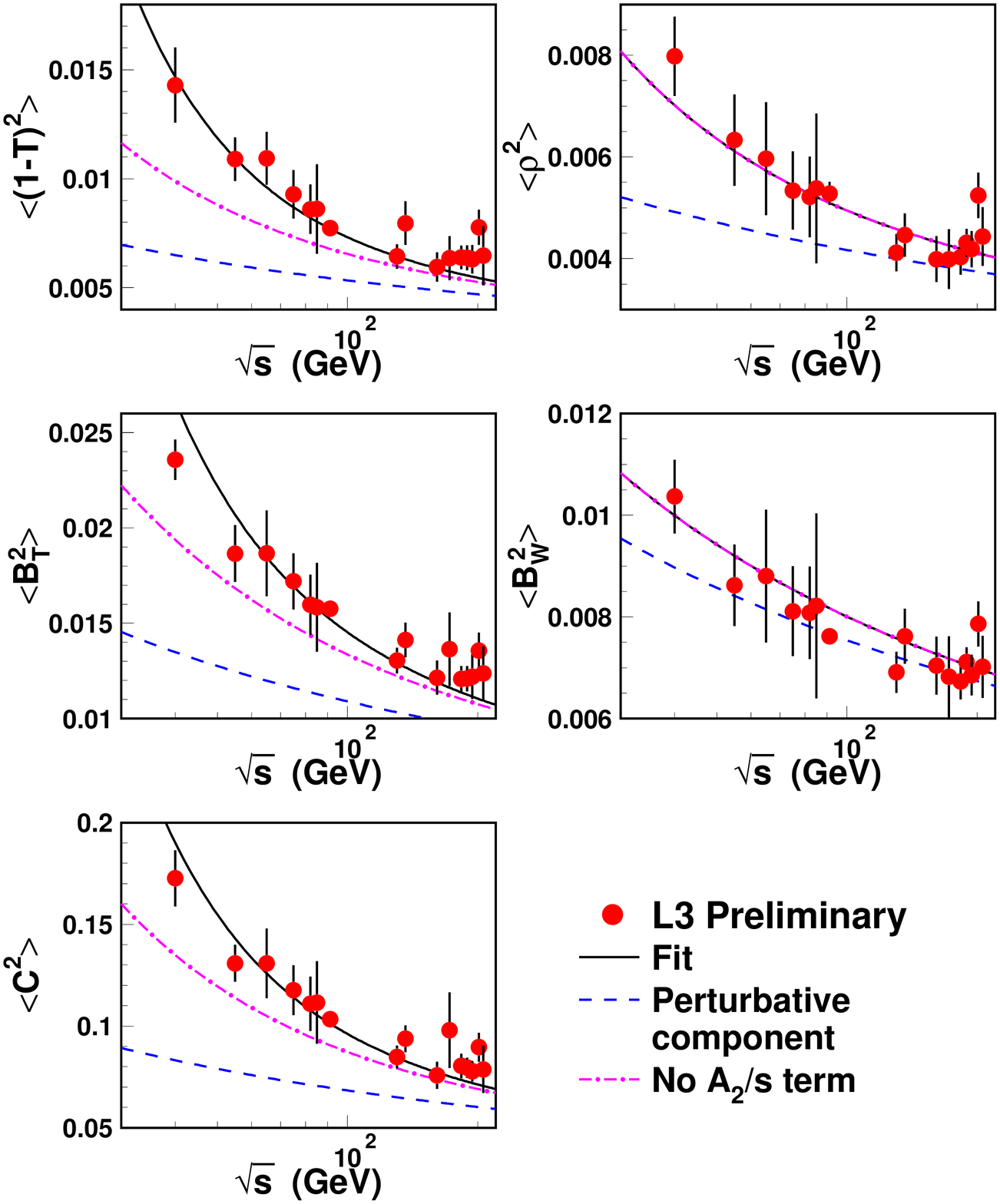}
\caption{Second moments of event shapes from L3~\cite{l3lep2data2}.}
\label{fig_l3}
\end{figure}

Figure~\ref{fig_l3} shows the results by L3 for 2nd moments from
analyses of the LEP~2 high energy data as well as from low energy data
obtained by selecting events at $\roots=\mz$ with hard initial or final
state photon radiation. The fits of the energy evolution of the 2nd
moments are based on a simple extension of the DMW model with $\langle
y^2\rangle=\langle y^2\rangle_{PT}+2\langle
y\rangle_{PT}PD_y+A_2/Q^2$. The quantity $A_2$ is fitted with \as\ and
\anull\ fixed to the results from fits to 1st moments and it is
found to be consistent with zero only for heavy jet mass and the wide
jet broadening. 

\subsection{Differential distributions}
\label{sec_dsdy}

\begin{figure}[!htb]
\vspace{9pt}
\includegraphics[width=\columnwidth]{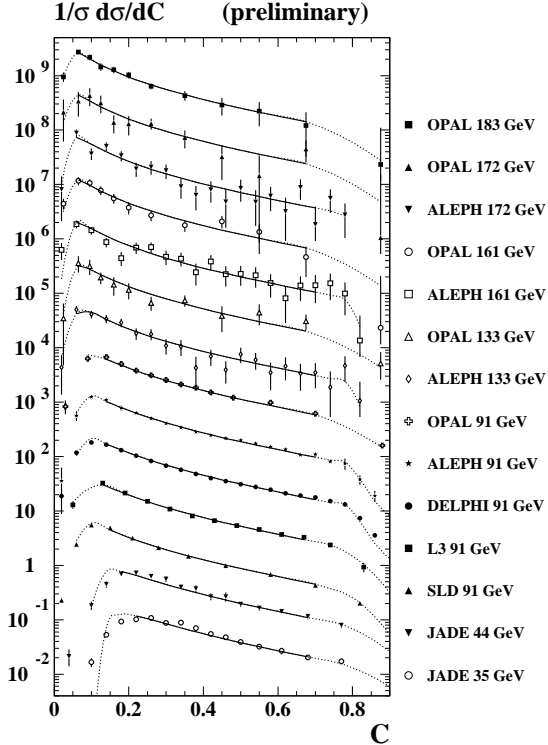}
\caption{Distributions of C-parameter~\cite{jadedist}.}
\label{fig_jade}
\end{figure}

In the DMW model power corrections to differential event shape
distributions are described by a shift $\sim1/Q$ of the perturbative
prediction. This prediction has been tested with reanalysed JADE data
together with data from SLD and the LEP experiments~\cite{jadedist}
and also with LEP data by
ALEPH~\cite{alephlep2data1}. Figure~\ref{fig_jade} presents
distributions of the C-parameter from $\roots=35$~GeV to 183~GeV with
a fit of the DMW prediction~\cite{jadedist}. The fitted prediction
within the fit ranges indicated by solid lines describes the data well
within the errors.

\section{STUDY OF QCD COLOUR FACTORS}

\begin{figure}[!htb]
\vspace{9pt}
\includegraphics[width=\columnwidth]{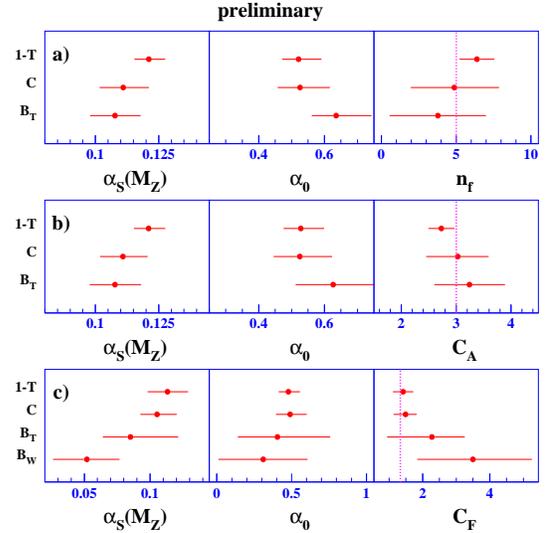}
\caption{Results for \asmz, \anull\ and the QCD colour factors from fits
to event shape distributions~\cite{colrun}.}
\label{fig_colrun}
\end{figure}

In this analysis fits of the DMW model to event shape distributions as
shown in section~\ref{sec_dsdy} are generalised to vary one of the QCD
colour factors \nf, \ca\ or \cf\ along with \as\ and
\anull~\cite{colrun}. Sensitivity to the colour factors comes mainly
from the running of \as\ in the perturbative prediction. Potential
biases from assuming standard QCD in the hadronisation corrections
should not be present. It is observed that fits to thrust and
C-parameter are stable while fits to total and wide jet broadening
don't converge well in all cases. Results of the fits to individual
colour factors, \as\ and \anull\ are shown in figure~\ref{fig_colrun}
and are consistent with standard QCD based on the SU(3) symmetry
group. The total uncertainties on the measurements of the colour
factors are competitive with traditional analyses using 4-jet final
states at $\roots=\mz$. 

\section{SUMMARY}

\begin{figure}[!htb]
\vspace{9pt}
\includegraphics[width=\columnwidth]{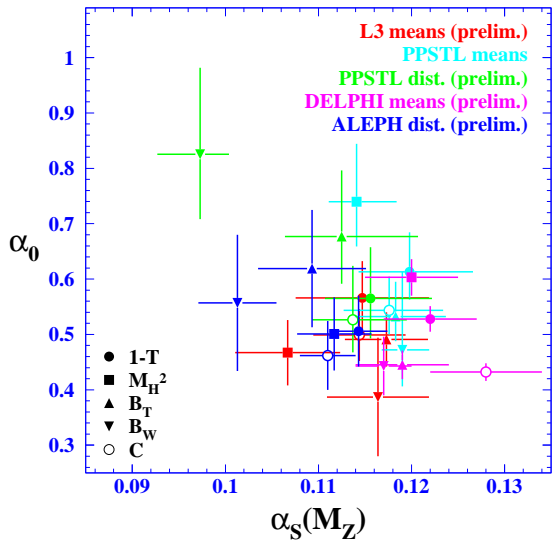}
\caption{Results for \as\ and \anull\ from analyses of the DMW power
correction model. }
\label{fig_summ}
\end{figure}

A summary of all results for \asmz\ and \anull\ of analyses based on the
DMW model is shown in figure~\ref{fig_summ}. The label PPSTL refers to
the analyses of data from PEP, PETRA, SLC, TRISTAN and LEP performed
together with reanalysed JADE data~\cite{jadenewas,jadec,jadedist}. The
results for \asmz\ are consistent with the current world average
$\asmz=0.118\pm0.003$~\cite{bethke00a} while the results for \anull\
mostly cluster within $\anull\simeq0.5\pm25\%$. The results for
\anull\ support the prediction of universality within 25\%. 

The other power corrections models are promising new approaches and
need experimental testing. From an experimental point of view it would
also be desirable to quantify the relations between the different
models. 

The study of QCD colour factors within the DMW model yields results
consistent with standard SU(3) QCD with thrust and C-parameter. The
total uncertainties on the colour factors are competitive with other
measurements of the colour factors. 

The field of power corrections in \epem\ annihilation is rapidly
developing and it gives one the opportunity to learn more about soft
QCD and the physics of hadronisation.


\section*{QUESTIONS}

\begin{description}
\item[M. Boutemeur, LMU Munich] Is it normal that your analysis [of
the QCD colour factors] works for both 1-T and C because they are
almost 100\% correlated? Do you know any other event shape variable
for which it works?
\item[Answer] It is consistent that the analysis works for \thr\ and \cp,
because they are correlated. However, the observables \bt\ and \bw\
are also highly correlated with \thr\ and \cp\ but the fits
are not always stable. This may indicate that the DMW model is not as
good an approximation for \bt\ and \bw\ as for \thr\ and \cp. Other
event shapes for which the analysis would be possible are heavy jet
mass and energy-energy-correlation. 
\item[Hasko Stenzel, MPI Munich] What are the indications of an
insufficient perturbative description of the wide jet broadening [in
the colour factor analysis]?
\item[Answer] The analysis of the colour factors used the latest
corrected calculations for the jet broadening observables. It is not
known if the perturbative or the power correction part of the
prediction for \bw\ is insufficient. 
\end{description}

\end{document}